\documentclass [a4paper,12pt] {article}

\usepackage{latexsym}
\usepackage{amsfonts}
\usepackage{amsmath}
\usepackage{amssymb}

  \def\pp{{\mathchoice
          {
              \kern 1pt%
              \raise 1pt
              \vbox{\hrule width5pt height0.4pt depth0pt
                    \kern -2pt
                    \hbox{\kern 2.3pt
                          \vrule width0.4pt height6pt depth0pt
                          }
                    \kern -2pt
                    \hrule width5pt height0.4pt depth0pt}%
                    \kern 1pt
           }
            {
              \kern 1pt%
              \raise 1pt
              \vbox{\hrule width4.3pt height0.4pt depth0pt
                    \kern -1.8pt
                    \hbox{\kern 1.95pt
                          \vrule width0.4pt height5.4pt depth0pt
                          }
                    \kern -1.8pt
                    \hrule width4.3pt height0.4pt depth0pt}%
                    \kern 1pt
            }
            {
              \kern 0.5pt%
              \raise 1pt
              \vbox{\hrule width4.0pt height0.3pt depth0pt
                    \kern -1.9pt  
                    \hbox{\kern 1.85pt
                          \vrule width0.3pt height5.7pt depth0pt
                          }
                    \kern -1.9pt
                    \hrule width4.0pt height0.3pt depth0pt}%
                    \kern 0.5pt
            }
            {
              \kern 0.5pt%
              \raise 1pt
              \vbox{\hrule width3.6pt height0.3pt depth0pt
                    \kern -1.5pt
                    \hbox{\kern 1.65pt
                          \vrule width0.3pt height4.5pt depth0pt
                          }
                    \kern -1.5pt
                    \hrule width3.6pt height0.3pt depth0pt}%
                    \kern 0.5pt
            }
        }}

  \def\mm{{\mathchoice
                       {
                             \kern 1pt
               \raise 1pt    \vbox{\hrule width5pt height0.4pt depth0pt
                                  \kern 2pt
                                  \hrule width5pt height0.4pt depth0pt}
                             \kern 1pt}
                       {
                            \kern 1pt
               \raise 1pt \vbox{\hrule width4.3pt height0.4pt depth0pt
                                  \kern 1.8pt
                                  \hrule width4.3pt height0.4pt depth0pt}
                             \kern 1pt}
                       {
                            \kern 0.5pt
               \raise 1pt
                            \vbox{\hrule width4.0pt height0.3pt depth0pt
                                  \kern 1.9pt
                                  \hrule width4.0pt height0.3pt depth0pt}
                            \kern 1pt}
                       {
                           \kern 0.5pt
             \raise 1pt  \vbox{\hrule width3.6pt height0.3pt depth0pt
                                  \kern 1.5pt
                                  \hrule width3.6pt height0.3pt depth0pt}
                           \kern 0.5pt}
                       }}

\catcode`@=11
\def\un#1{\relax\ifmmode\@@underline#1\else
        $\@@underline{\hbox{#1}}$\relax\fi}
\catcode`@=12

\let\du=\du                     

\def\a{\alpha}
\def\b{\beta}
\def\c{\chi}
\def\d{\delta}

\def\f{\phi}
\def\g{\gamma}

\def\k{\kappa}

\def\m{\mu}
\def\n{\nu}
\def\o{\omega}

\def\q{\theta}

\def\s{\sigma}

\def\x{\xi}

\def\F{\Phi}

\def\ve{\varepsilon}

\def\bo{{\raise-.5ex\hbox{\large$\Box$}}}               
\def\pa{\partial}                                       
\def\TH{{\raise.2ex\hbox{$\displaystyle \bigodot$}\mskip-4.7mu \llap H \;}}
\def\face{{\raise.2ex\hbox{$\displaystyle \bigodot$}\mskip-2.2mu \llap {$\ddot
        \smile$}}}                                      

\def\Tilde#1{\widetilde{#1}}                    
\def\Bar#1{\overline{#1}}                       
\def\abs#1{\left| #1\right|}                    
\def\leftrightarrowfill{$\mathsurround=0pt \mathord\leftarrow \mkern-6mu
        \cleaders\hbox{$\mkern-2mu \mathord- \mkern-2mu$}\hfill
        \mkern-6mu \mathord\rightarrow$}
\def\dvec#1{\vbox{\ialign{##\crcr
        \leftrightarrowfill\crcr\noalign{\kern-1pt\nointerlineskip}
        $\hfil\displaystyle{#1}\hfil$\crcr}}}           
\def\dt#1{{\buildrel {\hbox{\LARGE .}} \over {#1}}}     

\def\frac#1#2{{\textstyle{#1\over\vphantom2\smash{\raise.20ex
        \hbox{$\scriptstyle{#2}$}}}}}                   
\def\sfrac#1#2{{\vphantom1\smash{\lower.5ex\hbox{\small$#1$}}\over
        \vphantom1\smash{\raise.4ex\hbox{\small$#2$}}}} 
\def\bfrac#1#2{{\vphantom1\smash{\lower.5ex\hbox{$#1$}}\over
        \vphantom1\smash{\raise.3ex\hbox{$#2$}}}}       
\def\afrac#1#2{{\vphantom1\smash{\lower.5ex\hbox{$#1$}}\over#2}}    
\def\on#1#2{\mathop{\null#2}\limits^{#1}}               
\def\bvec#1{\on\leftarrow{#1}}                  

\def\[{\lfloor{\hskip 0.35pt}\!\!\!\lceil}
\def\]{\rfloor{\hskip 0.35pt}\!\!\!\rceil}

\def\du#1#2{_{#1}{}^{#2}}

\def\fracm#1#2{\hbox{\large{${\frac{{#1}}{{#2}}}$}}}
\def\ha{{\fracmm12}}

\def\un{\underline}
\def\fracmm#1#2{{{#1}\over{#2}}}

\def\low#1{{\raise -3pt\hbox{${\hskip 0.75pt}\!_{#1}$}}}

\def\Dot#1{\buildrel{_{_{\hskip 0.01in}\bullet}}\over{#1}}
\def\dt#1{\Dot{#1}}

\def\Tilde#1{{\widetilde{#1}}\hskip 0.015in}

\newskip\humongous \humongous=0pt plus 1000pt minus 1000pt
\def\caja{\mathsurround=0pt}
\def\eqalign#1{\,\vcenter{\openup2\jot \caja
        \ialign{\strut \hfil$\displaystyle{##}$&$
        \displaystyle{{}##}$\hfil\crcr#1\crcr}}\,}
\newif\ifdtup

\parskip=\medskipamount                 
\thicklines                         

\begin{document}
\thispagestyle{empty}

{\hbox to\hsize{
\vbox{\noindent June 2005 \hfill hep-th/0506071}}}

\noindent
\vskip1.3cm
\begin{center}

{\Large\bf $N=\fracm{1}{2}$  supersymmetric four-dimensional non-linear
\vglue.1in
           $\s$-models from non-anti-commutative superspace~\footnote{
Supported in part by the Japanese Society for Promotion of Science (JSPS)}}
\vglue.2in

Tomoya Hatanaka~\footnote{Email address: thata@kiso.phys.metro-u.ac.jp},
Sergei V. Ketov~\footnote{Email address: ketov@phys.metro-u.ac.jp},
Yoshishige Kobayashi~\footnote{Email address: yosh@phys.metro-u.ac.jp},
and Shin Sasaki~\footnote{Email address: shin-s@phys.metro-u.ac.jp}

{\it Department of Physics\\
     Tokyo Metropolitan University\\
     1--1 Minami-osawa, Hachioji-shi\\
     Tokyo 192--0397, Japan}
\end{center}
\vglue.2in
\begin{center}
{\Large\bf Abstract}
\end{center}

The component structure of a generic N=1/2 supersymmetric Non-Linear 
Sigma-Model (NLSM) defined in the four-dimensional (Euclidean) 
Non-Anti-Commutative (NAC) superspace is investigated in detail. The most 
general NLSM is described in terms of arbitrary K\"ahler potential, and chiral 
and anti-chiral superpotentials. The case of a single chiral superfield gives 
rise to splitting of the NLSM potentials, whereas the case of several chiral 
superfields results in smearing (or fuzziness) of the NLSM potentials, while
both effects are controlled by the auxiliary fields. We eliminate the 
auxiliary fields by solving their algebraic equations of motion, and 
demonstrate that the results are dependent upon whether the auxiliary 
integrations responsible for the fuzziness are performed before or after 
elimination of the auxiliary fields. There is no ambiguity in the case of 
splitting, i.e. for a single chiral superfield. Fully explicit results are 
derived in the case of the N=1/2 supersymmetric NAC-deformed $CP^n$ NLSM in 
four dimensions. Here we find another surprise that our results differ from 
the N=1/2 supersymmetric $CP^n$ NLSM derived by the quotient construction from
the N=1/2 supersymmetric NAC-deformed gauge theory. We conclude that an N=1/2 
supersymmetric deformation of a generic NLSM from the NAC superspace is not 
unique.
 
\newpage

\section{Introduction}

Investigation of various aspects of Non-Anti-Commutative (NAC) superspace and 
related deformations of supersymmetric field theories are the subject of
intensive studies during the last two years (see, e.g., 
refs.~\cite{sei,chandra,chuo,buch,nac1,spain,nbi,nac2} directly related to our 
title, and the references therein for the earlier work in the NAC-deformed N=1 
superspace). Our general motivation is to enhance understanding of the role of 
spacetime in supersymmetry, while keeping globally supersymmetric field theory
under control. Yet additional motivation is provided by string theory, where
the non-anti-commutativity can be related to a constant (self-dual) 
gravi-photon background (see e.g., ref.~\cite{sei} and references therein).  

This paper is devoted to the NAC-deformed supersymmetric Non-Linear 
Sigma-Models (NLSM) in four dimensions. The bosonic NLSM we consider are the 
most general scalar field theories, {\it without\/} any gauge fields or higher
derivatives. Their (undeformed) N=1 supersymmetric extension is well known to
require K\"ahler geometry of the NLSM metric and a holomorphic scalar 
superpotential (see e.g., the pioneering paper \cite{zumino} and 
ref.~\cite{nlsm} for a review).

We work in four-dimensional Euclidean~\footnote{The use of 
Atiyah-Ward spacetime of the signature $(+,+,-,-)$ is another possibility
\cite{kgn}.} N=1 superspace
$(x^{\m},\q^{\a},\bar{\q}^{\dt{\a}})$, and use the standard notation \cite{wb}.
The NAC deformation is given by  
$$ \{ \q^{\a},\q^{\b} \}_*=C^{\a\b}~~,
\eqno(1.1)$$
where $C^{\a\b}$ are some constants. The remaining superspace coordinates in 
the chiral basis ($y^{\m}=x^{\m}+i\q\s^{\m}\bar{\q}$, $\m,\n=1,2,3,4$ and 
$\a,\b,\ldots=1,2$) still (anti)commute,
$$\[y^{\m},y^{\n}\]= \{ \bar{\q}^{\dt{\a}},\bar{\q}^{\dt{\b}} \}=
\{ \q^{\a},\bar{\q}^{\dt{\b}} \}=\[y^{\m},\q^{\a}\]=
\[y^{\m},\bar{\q}^{\dt{\a}}\]=0~.\eqno(1.2)$$

The $C^{\a\b}\neq 0$ explicitly break the four-dimensional `Lorentz' 
invariance at the fundamental level. The NAC nature of $\q$'s can be fully 
taken into account by using the Moyal-Weyl-type (star) product of
superfields \cite{sei}~,
 $$ f(\q)* g(\q)=f(\q)\,
\exp\left(-\fracmm{C^{\a\b}}{2}\fracmm{\bvec{\pa}}{\pa
\q^{\a}}\fracmm{\vec{\pa}}{\pa\q^{\b}}\right)g(\q)~, \eqno(1.3)$$
which respects the N=1 superspace chirality. The star product (1.3) is 
polynomial in the deformation parameter~,
$$ f(\q)*g(\q)=fg +(-1)^{{\rm deg}f}\fracmm{C^{\a\b}}{2}
\fracmm{\pa f}{\pa\q^{\a}}\fracmm{\pa g}{\pa\q^{\b}}-\det\,C
\fracmm{\pa^2 f}{\pa\q^2}\fracmm{\pa^2 g}{\pa\q^2}~~,\eqno(1.4)$$
where we have used the identity
$$ \det C = \fracm{1}{2}\ve_{\a\g}\ve_{\b\d}C^{\a\b}C^{\g\d}~,\eqno(1.5)$$
and the notation
$$ \fracmm{\pa^2}{\pa\q^2}= \fracm{1}{4}\ve^{\a\b}\fracmm{\pa}{\pa\q^{\a}}
\fracmm{\pa}{\pa\q^{\b}}~~.\eqno(1.6)$$

We also use the following book-keeping notation for 2-component spinors:
$$ \q\c=\q^{\a}\c_{\a}~,\quad  \bar{\q}\bar{\c}=\bar{\q}_{\dt{\a}}
\bar{\c}^{\dt{\a}}~,\quad \q^2= \q^{\a}\q_{\a}~,\quad 
\bar{\q}^2= \bar{\q}_{\dt{\a}}\bar{\q}^{\dt{\a}}.\eqno(1.7)$$
The spinorial indices are raised and lowered by the use of two-dimensional 
Levi-Civita symbols \cite{wb}. Grassmann integration amounts to Grassmann 
differentiation. The anti-chiral covariant derivative in the chiral superspace
basis is $\bar{D}_{\dt{\a}}=-\pa/ \pa \bar{\q}^{\dt{\a}}$. The field component 
expansion of a chiral superfield $\F$ reads 
$$ \F(y,\q)= \f(y) +\sqrt{2}\q\c(y)+\q^2 M(y)~~.\eqno(1.8)$$
An anti-chiral superfield $\Bar{\F}$ in the chiral basis is given by
$$\eqalign{
\Bar{\F}(y^{\m}-2i\q\s^{\m}\bar{\q},\bar{\q})=  ~ & ~ 
\bar{\f}(y) + \sqrt{2}\bar{\q}\bar{\c}(y) +\bar{\q}^2\bar{M}(y)  \cr
 ~ & ~ +\sqrt{2}\q\left( i\s^{\m}\pa_{\m}\bar{\c}(y)\bar{\q}^2-i\sqrt{2}\s^{\m}
\bar{\q}\pa_{\m}\bar{\f}(y)\right)+\q^2\bar{\q}^2\bo\bar{\f}(y)~,\cr}
\eqno(1.9)$$
where $\bo =\pa_{\m}\pa_{\m}$. The bars over fields serve to distinguish 
between the `left' and `right' components that are truly independent in 
Euclidean spacetime.

The non-anticommutativity $C_{\a\b}\neq 0$ also explicitly breaks {\it half} 
of the original N=1 supersymmetry \cite{sei}. Only the chiral subalgebra 
generated by the chiral supercharges (in the chiral basis) 
$Q_{\a}=\pa/\pa\q^{\a}$ is preserved, with $\{ Q_{\a},Q_{\b} \}_*=0$, thus
defining what is now called N=1/2 supersymmetry. The use of the NAC-deformed
superspace allows one to keep N=1/2 supersymmetry manifest. The N=1/2
supersymmetry transformation laws of the chiral and anti-chiral superfield
components in eqs.~(1.8) and (1.9) are as follows:
$$\d\f=\sqrt{2}\ve^{\a}\c_{\a}~~,\quad \d\c_{\a}=\sqrt{2}\ve_{\a}M~,\quad
\d M =0~,\eqno(1.10)$$
and
$$\d\bar{\f}=0~,\quad \d\bar{\c}^{\dt{\a}}=-i\sqrt{2}
(\tilde{\s}_{\m})^{\dt{\a}\b}\ve_{\b}\pa_{\m}\bar{\f}~,\quad
\d\bar{M}=-i\sqrt{2}\pa_{\m}\bar{\c}_{\dt{\a}}(\tilde{\s}_{\m})^{\dt{\a}\b}
\ve_{\b}~,\eqno(1.11)$$
respectively, where we have introduced the N=1/2 supersymmetry (chiral) 
parameter $\ve^{\a}$.

The most general four-dimensional N=1 supersymmetric NLSM action (without any
gauge fields) is given by
$$ S[\F,\Bar{\F}]
 = \int d^4 y \left[ \int d^2\q d^2\bar{\q}\, K(\F^i,\Bar{\F}{}^{\bar{j}})+
\int d^2\q\, W(\F^i) + \int  d^2\bar{\q}\,\Bar{W}(\Bar{\F}{}^{\bar{j}})
\right]~.\eqno(1.12)$$
This action is completely specified by the K\"ahler superpotential 
$K(\F,\Bar{\F})$, the scalar superpotential $W(\F)$, and the anti-chiral 
superpotential $\Bar{W}(\Bar{\F})$, in terms of some number $n$  of 
chiral and anti-chiral superfields, $i,\bar{j}=1,2,\ldots,n$. In Euclidean 
superspace the functions $W(\F)$ and $\Bar{W}(\Bar{\F})$ are independent upon 
each other.

The problem of computing the NAC-deformed extension of eq.~(1.12) in four 
dimensions, after a `Seiberg-Witten map' (i.e. after evaluating all the
star products), was solved in ref.~\cite{buch} in the case of a single 
(anti)chiral superfield. That solution was extended to the case of several 
(anti)chiral superfields in ref.~\cite{nbi}. In both cases the perturbative 
solutions were found, i.e. in terms of the infinite sums with respect to the
deformation parameter and the auxiliary fields. A similar problem in two 
dimensions was perturbatively solved in ref.~\cite{chandra}. A non-perturbative
solution (i.e. in a closed form, in terms of finite functions) was found in 
ref.~\cite{spain} in two dimensions, and in ref.~\cite{nac2} in four 
dimensions. The calculations behind each result appear to be quite extensive, 
so it is non-trivial to make a comparison. In our opinion, an explicit 
summation is necessary anyway, both for comparison and elimination of the 
auxiliary fields. Getting the results in a closed form is also crucial for any
physical applications of the NAC-deformation and its geometrical 
interpretation, as well as in investigating concrete examples (see below). In 
this paper we use our four-dimensional  results \cite{nac2} that are in 
precise agreement with the basis formulae of ref.~\cite{spain}, after 
dimensional reduction to two dimensions. The non-perturbative results of 
refs.~\cite{spain} and \cite{nac2} presumably amount to the full summation of 
the perturbative results in refs.~\cite{chandra} and \cite{buch,nbi}, 
respectively. 
  
We use the chiral basis that is most suitable for computing NAC-deformation, 
by reducing the most non-trivial problem of calculation of the NAC-deformed 
K\"ahler superpotential to that for the NAC-deformed scalar 
superpotential \cite{nac2}. The simple general results about the 
NAC-deformed scalar superpotentials were found in refs.~\cite{nac1,spain}. 
Our major concern in this paper is getting solutions to the auxiliary fields
that control splitting or fuzziness in the NLSM target space.

Our paper is organized as follows. In sect.~2 we summarize the earlier results
\cite{nac1,nac2} that represent our setup here. In sect.~3 we eliminate the 
auxiliary fields in a generic NAC-deformed NLSM, by using implicit functions.
The case of a single (anti)chiral superfield is also considered separately. In
 sect.~4 we specify our results to the case of the $CP^n$ NLSM. The $CP^1$ 
case is fully investigated. Sect.~5 is our conclusion.

\section{The NAC K\"ahler potential and superpotentials}

We use the following notation valid for any function $F(\f,\bar{\f})$:
$$ F_{,i_1i_2\cdots i_s \bar{p}_1\bar{p}_2\cdots \bar{p}_t}=
\fracmm{\pa^{s+t}F}{\pa\f^{i_1}\pa\f^{i_2}\cdots\f^{i_s}
\pa\bar{\f}^{\bar{p}_1}\pa\bar{\f}^{\bar{p}_2}\cdots\pa
\bar{\f}^{\bar{p}_t}}~~~~,\eqno(2.1)$$
and the Grassmann integral normalisation $ \int d^2\q\,\q^2=1$. The actual
deformation parameter, in the case of the NAC-deformed field theory (1.12), 
appears to be 
$$ c=\sqrt{-\det C}~,\eqno(2.2)$$
where we have used the definition \cite{sei}
$$  \det C = \fracm{1}{2}\ve_{\a\g}\ve_{\b\d}C^{\a\b}C^{\g\d}~~.\eqno(2.3)$$
As a result, unlike the case of the NAC-deformed supersymmetric gauge theories
 \cite{sei}, the NAC-deformation of the NLSM field theory (1.12) appears to be
 `Lorentz'-invariant \cite{nac1,nac2}. 

A simple non-perturbative formula, describing an arbitrary NAC-deformed 
scalar superpotential $V$ depending upon a single chiral superfield $\F$, was 
found in ref.~\cite{nac1},
$$ \int d^2\q\,V_*(\F)=\fracm{1}{2c}\left[ V(\f+cM)-V(\f-cM)\right]
-\fracmm{\c^2}{4cM}\left[ V_{,\f}(\f+cM)- V_{,\f}(\f-cM)\right]~.\eqno(2.4)$$
The NAC-deformation in the single superfield case thus gives rise to the split
of the scalar potential, which is controlled by the auxiliary field $M$. When 
using an elementary identity
$$ f(x+a)-f(x-a) =a\fracmm{\pa}{\pa x}
\int^{+1}_{-1} d\x\,f(x+\x a)~,\eqno(2.5)$$
valid for any function $f$, we can rewrite eq.~(2.4) to the equivalent form,
as in ref.~\cite{spain},
$$ \int d^2\q\,V_*(\F)=\fracm{1}{2}M\fracmm{\pa}{\pa\f}\int^{+1}_{-1} d\x\,
V(\f+\x cM) -\fracmm{1}{4}\c^2\fracmm{\pa^2}{\pa\f^2}\int^{+1}_{-1} d\x\,
V(\f+\x cM)~,\eqno(2.6)$$    
Similarly, in the case of several chiral superfields, one finds \cite{spain}
$$ \int d^2\q\,V_*(\F^I)=\fracm{1}{2}M^I\fracmm{\pa}{\pa\f^I}\Tilde{V}(\f,M)
-\fracmm{1}{4}(\c^I\c^J)\fracmm{\pa^2}{\pa\f^I\pa\f^J}\Tilde{V}(\f,M)~,
\eqno(2.7)$$   
in terms of the auxiliary pre-potential $\Tilde{V}$ \cite{spain},
$$\Tilde{V}(\f,M)=\int^{+1}_{-1} d\x\,V(\f^I+\x cM^I)~.\eqno(2.8)$$
Hence the NAC-deformation of a generic scalar superpotential $V$ results
in its smearing or fuzziness controlled by the auxiliary fields $M^I$ 
\cite{spain}.

A calculation of the NAC deformed K\"ahler potential 
$$ \int d^4y\,L_{\rm kin.}\equiv\int d^4y\int d^2\q d^2\bar{\q}\, 
K(\F^i,\Bar{\F}{}^{\bar{j}})_* \eqno(2.9)$$
can be reduced to eqs.~(2.4) or (2.7), when using a chiral
reduction in superspace, with the following result \cite{nac2}:
$$ \eqalign{
L_{\rm kin.}~ = ~& ~ \fracm{1}{2} M^iY_{,i}+
\fracm{1}{2}\pa^{\m}\bar{\f}{}^{\bar{p}}\pa_{\m}\bar{\f}{}^{\bar{q}}
Z_{,\bar{p}\bar{q}} + \fracm{1}{2}\bo\bar{\f}{}^{\bar{p}}
Z_{,\bar{p}} -\fracm{1}{4}(\c^i\c^j)Y_{,ij} \cr
 ~& ~ -\fracm{1}{2}i(\c^i\s^{\m}\bar{\c}{}^{\bar{p}})\pa_{\m}
\bar{\f}{}^{\bar{q}}Z_{,i\bar{p}\bar{q}}
-\fracm{1}{2}i(\c^i\s^{\m}\pa_{\m}\bar{\c}{}^{\bar{p}})Z_{,i\bar{p}}~,\cr}
\eqno(2.10)$$
where we have introduced the (component) smeared K\"ahler pre-potential 
$$ Z(\f,\bar{\f},M)=\int^{+1}_{-1}d\x\, K^{\x} \quad {\rm with}\quad
K^{\x}\equiv K(\f^i+\x cM^i,\bar{\f}{}^{\bar{j}})~~,\eqno(2.11)$$
as well as the extra (auxiliary) pre-potential 
$$ Y(\f,\bar{\f},M,\bar{M})=\bar{M}{}^{\bar{p}}Z_{,\bar{p}}
 -\fracm{1}{2}(\bar{\c}{}^{\bar{p}}\bar{\c}{}^{\bar{q}})
Z_{,\bar{p}\bar{q}} +c\int^{+1}_{-1}d\x \x\left[ 
\pa^{\m}\bar{\f}{}^{\bar{p}}\pa_{\m}\bar{\f}{}^{\bar{q}}
K^{\x}_{,\bar{p}\bar{q}} +\bo\bar{\f}{}^{\bar{p}}K^{\x}_{,\bar{p}}\right]~,
\eqno(2.12)$$
as in ref.~\cite{spain}. It is not difficult to check that eq.~(2.10) does 
reduce to the standard (K\"ahler) N=1 supersymmetric NLSM \cite{zumino} in 
the limit $c\to 0$. Also, in the case of a free (bilinear) K\"ahler potential 
$K=\d_{i\bar{j}}\F^i\bar{\F}^{\bar{j}}$, there is no deformation at all. 

The NAC-deformed scalar superpotentials $W(\F)_*$ and $\bar{W}(\bar{\F})_*$ 
imply, via eqs.~(2.7) and (2.8), that the following component terms are to be 
added to eq.~(2.10):
$$ L_{\rm pot.}=\fracm{1}{2}M^i\Tilde{W}_{,i}-\fracm{1}{4}(\c^i\c^j)
\Tilde{W}_{,ij}+ \bar{M}^{\bar{p}}\bar{W}_{,\bar{p}}-  
\fracm{1}{2}(\bar{\c}^{\bar{p}}\bar{\c}^{\bar{q}})\bar{W}_{,\bar{p}\bar{q}}~.
\eqno(2.13)$$
where we have introduced the smeared scalar pre-potential \cite{spain}
$$ \Tilde{W}(\f,M)= \int^{+1}_{-1}d\x\, W(\f^i+\x cM^i)~~.\eqno(2.14)$$
The anti-chiral scalar superpotential terms are {\it inert} under the 
NAC-deformation \cite{buch,nac1,spain,nbi,nac2}.

The $\x$-integrations in the equations above represent the smearing effects. 
However, the smearing is merely apparent in the case of a single chiral 
superfield, which gives rise to the splitting (2.4) only. This can also be
directly demonstrated from eq.~(2.10) when using the identity (2.5) together
with the related identity \cite{nac2}
$$ f(x+a) + f(x-a) =\int^{+1}_{-1}d\x\,f(x+\x a) + a\fracmm{\pa}{\pa x}
\int^{+1}_{-1}d\x\,\x f(x+\x a)~.\eqno(2.15)$$ 
The single superfield case thus appears to be special, so that a sum of 
eq.~(2.10) and (2.13) can be rewritten to the bosonic contribution \cite{nac2}
$$\eqalign{
L_{\rm bos.} = & 
+\fracm{1}{2}\pa^{\m}\bar{\f}\pa_{\m}\bar{\f}\left[
K_{,\bar{\f}\bar{\f}}(\f+cM,\bar{\f})+K_{,\bar{\f}\bar{\f}}(\f-cM,\bar{\f})
\right] \cr
& + \fracm{1}{2}\bo\bar{\f}\left[
K_{,\bar{\f}}(\f+cM,\bar{\f})+K_{,\bar{\f}}(\f-cM,\bar{\f})\right] \cr
& + \fracmm{\bar{M}}{2c}\left[ K_{,\bar{\f}}(\f+cM,\bar{\f}) 
 -K_{,\bar{\f}}(\f-cM,\bar{\f}) \right] \cr
& + \fracmm{1}{2c}\left[ W(\f+cM) -W(\f-cM)\right]+\bar{M}
\fracmm{\pa\bar{W}}{\pa\bar{\f}}~~,\cr} \eqno(2.16)$$
supplemented by the following fermionic terms \cite{nac2}:
$$\eqalign{
L_{\rm ferm.} = & -\fracmm{1}{4c}\bar{\c}^2\left[ 
K_{,\bar{\f}\bar{\f}}(\f+cM,\bar{\f})-K_{,\bar{\f}\bar{\f}}(\f-cM,\bar{\f})
\right] \cr
& -\fracmm{i}{2cM}(\c\s^{\m}\bar{\c})\pa_{\m}\bar{\f}\left[
K_{,\bar{\f}\bar{\f}}(\f+cM,\bar{\f})-K_{,\bar{\f}\bar{\f}}(\f-cM,\bar{\f})
\right] \cr
& -\fracmm{i}{2cM}(\c\s^{\m}\pa_{\m}\bar{\c})\left[
K_{,\bar{\f}}(\f+cM,\bar{\f})-K_{,\bar{\f}}(\f-cM,\bar{\f})
\right] \cr
& -\fracmm{\bar{M}}{4cM}\c^2 
\left[ K_{,\f\bar{\f}}(\f+cM,\bar{\f})-K_{,\f\bar{\f}}(\f-cM,\bar{\f})
\right]  \cr
& -\fracmm{1}{4M}\c^2 \pa^{\m}\bar{\f}\pa_{\m}\bar{\f} 
\left[ K_{,\f\bar{\f}\bar{\f}}(\f+cM,\bar{\f})+K_{,\f\bar{\f}\bar{\f}}
(\f-cM,\bar{\f})\right]  \cr
& +\fracmm{1}{4cM^2}\c^2 \pa^{\m}\bar{\f}\pa_{\m}\bar{\f} 
\left[ K_{,\bar{\f}\bar{\f}}(\f+cM,\bar{\f})-K_{,\bar{\f}\bar{\f}}
(\f-cM,\bar{\f})\right]  \cr
& -\fracmm{1}{4M}\c^2\bo\bar{\f}
\left[ K_{,\f\bar{\f}}(\f+cM,\bar{\f})+K_{,\f\bar{\f}}
(\f-cM,\bar{\f})\right]  \cr 
& +\fracmm{1}{4cM^2}\c^2\bo\bar{\f}
\left[ K_{,\bar{\f}}(\f+cM,\bar{\f})-K_{,\bar{\f}}
(\f-cM,\bar{\f})\right]  \cr 
& +\fracmm{1}{8cM}\c^2\bar{\c}^2
\left[ K_{,\f\bar{\f}\bar{\f}}(\f+cM,\bar{\f})-K_{,\f\bar{\f}\bar{\f}}
(\f-cM,\bar{\f})\right]  \cr\
& -\fracmm{1}{4cM}\c^2\left[ W_{,\f}(\f+cM) - W_{,\f}(\f-cM)\right]
-\fracm{1}{2}\bar{\c}^2 \bar{W}_{,\bar{\f}\bar{\f}}~.\cr}\eqno(2.17)$$

The anti-chiral auxiliary fields $\bar{M}^{\bar{p}}$ enter the action (2.10)
linearly (as Lagrange multipliers), while their algebraic equations of motion,
$$\fracm{1}{2}M^i Z_{,i\bar{p}} - \fracm{1}{4}(\c^i\c^j)Z_{,ij\bar{p}}
+\bar{W}_{,\bar{p}}=0~, \eqno(2.18)$$ 
represent the non-linear set of equations on the auxiliary fields 
$M^i=M^i(\f,\bar{\f})$.~\footnote{Equation (2.18) is not a linear system 
because the function $Z$ is $M$-dependent --- see eq.~(2.11).} As a result, 
the bosonic scalar potential in components is given by \cite{nac2} 
$$V_{\rm scalar}(\f,\bar{\f}) = 
\left. \fracm{1}{2}M^i\Tilde{W}_{,i}\right|_{M=M(\f,\bar{\f})}~~.\eqno(2.19)$$
	
\section{Solving for the auxiliary fields in a generic case}

The NAC-deformed NLSM in sect.~2 is completely specified by a K\"ahler function
$K(\F,\bar{\F})$, a chiral function $W(\F)$, an anti-chiral function 
$\bar{W}(\bar{\F})$ and a constant deformation parameter $c$. As a matter of
fact, we didn't really used the constancy of $c$, so our results in sect.~2
are still valid even for a coordinate-dependent NAC deformation $c(y)$. The 
very possibility of such extension preserving N=1/2 supersymmetry was 
conjectured in ref.~\cite{shif} and further investigated in ref.~\cite{arg}.
In this section we assume, however, that $c=const$ for simplicity. 

A NAC-deformation is only possible in Euclidean spacetime.  However, we may 
try to press further, by analytically continue our results into Minkowski 
spacetime (this would require $\det C < 0$). Then the natural place for 
possible physical applications of NAC-deformation could be given by confining 
supersymmetric gauge field theories whose low-energy (IR) effective action 
takes the form of eq.~(1.12) in terms some colorless composite scalar 
superfields $\F$ and $\bar{\F}$ known as the glueball superfields. The use of 
Konishi anomaly equations \cite{konishi} allows one to construct exact 
effective superpotentials of the glueball superfields in various N=1 
supersymmetric quantum gauge theories with some number of colors and flavors 
(see e.g., ref.~\cite{five} for details), generalizing the standard 
Veneziano-Yankielowicz effective potential \cite{vy}. This may provide
some natural input for the NAC deformation presumably describing some 
supergravitational corrections to the low-energy effective field theory, as
well as for a possible dynamical supersymmetry breaking (see ref.~\cite{nac1} 
for some attempts in this direction).~\footnote{The Konishi anomaly in N=1/2 
supersymmetry was recently studied in ref.~\cite{chu}.} From that physical 
point of view, the fact that the  NAC-deformation gives rise to 
a `Lorentz'-invariant action is clearly a positive feature, whereas the 
apparent non-Hermiticity of the NAC-deformed action is clearly a negative 
feature. In this paper we are going to keep the chiral and  anti-chiral 
functions to be arbitrary and concentrate on an investigation of the
NAC-deformed kinetic terms.

Solving for the auxiliary fields in eq.~(2.10) represents not only a technical
but also a conceptual problem because of the smearing effects described by the
$\x$-integrations. To bring the kinetic terms in eqs.~(2.10) and (2.16) to the 
standard NLSM form (i.e. without the second order spacial derivatives), one 
has to integrate by parts that leads to the appearance of the spacial 
derivatives of the auxiliary fields. This implies that one has to solve 
eq.~(2.18) {\it before} integration by parts. Let $M^i=M^i(\f,\bar{\f})$ be
a solution to eq.~(2.18), and let's ignore fermions for simplicity 
$(\c^i_{\a}=\bar{\c}_{\dt{\a}}^{\bar{p}}=0)$. Substituting the solution back 
to the Lagrangian (2.10) and integrating by parts yields
$$\eqalign{
L_{\rm kin.} (\f,\bar{\f})= & 
-\fracm{1}{2}(\pa_{\m}\bar{\f}^{\bar{p}}\pa_{\m}\f^q)\int^{+1}_{-1}d\x\left[
K^{\x}_{,\bar{p}q}+2c\x M^i_{,q}K^{\x}_{,\bar{p}i}+c\x M^iK^{\x}_{,\bar{p}iq}
+c^2\x^2M^iK^{\x}_{,\bar{p}ij}M^j_{,q}\right] \cr
& -\fracm{1}{2}(\pa_{\m}\bar{\f}^{\bar{p}}\pa_{\m}\bar{\f}^{\bar{q}})
\int^{+1}_{-1}d\x\left[ 2c\x K^{\x}_{,\bar{p}i}M^i_{,\bar{q}}+c^2\x^2M^i
K^{\x}_{,\bar{p}ij}M^j_{,\bar{q}}\right]~~.\cr}\eqno(3.1)$$
It is now apparent that the NAC-deformation does not preserve the original 
K\"ahler geometry in eq.~(1.12), though the absence of $(\pa_{\m}\f)^2$ terms
and the particular structure of various contributions to eq.~(3.1) are
remarkable. The action (3.1) takes the form of a generic NLSM, being merely 
dependent upon mixed derivatives of the K\"ahler function, so that the 
original K\"ahler gauge invariance of eq.~(1.12),
$$ K(\f, \bar{\f})\to K(\f, \bar{\f})+f(\f) + \bar{f}(\bar{\f})~,\eqno(3.2)$$
with arbitrary gauge functions $f(\f)$ and $\bar{f}(\bar{\f})$ is still 
preserved.

However, a problem arises with the $\x$-integrations in eq.~(3.1), because the
fields $M^i$ are no longer independent and, hence, eq.~(2.5) cannot be applied.
Instead, eq.~(2.5) may have to be replaced by a more general identity
$$\eqalign{
\fracmm{d}{dx}\int^{+1}_{-1} d\x\, f(x + \x a(x)) = 
& \fracmm{1}{a(x)}\left[ f(x+a(x)) -f(x-a(x))\right] \cr
& +\fracmm{1}{a(x)}\left[ f(x+a(x)) + f(x-a(x))\right] \fracmm{da}{dx} \cr
& -\fracmm{1}{a^2(x)}\left[ F(x+a(x)) -F(x-a(x))\right] \fracmm{da}{dx}~, \cr}
\eqno(3.3)$$
where we have defined $F(x)=\int^x f(\x)d\x$. We believe, however, that this 
way of doing is not quite correct from the viewpoint of the original star 
product that has nothing to do with dynamics. In other words, the elimination
of the auxiliary fields and the $\x$-integrations do not commute, while, in
our opinion, all the $\x$-integrations have to be done {\it before} solving 
for the auxiliary fields. This procedure will be adopted in our explicit 
calculations for the $CP^n$ case in the next sect.~4. It is worth mentioning
that the ordering problem does not arise in the case of a sinlge chiral 
superfield, because there is no need for $\x$-integrations there (i.e. when 
smearing is replaced by splitting). In the last case, when using eq.~(2.16), 
we find
$$ \eqalign{
L_{\rm kin.} (\f,\bar{\f})= & 
-\fracm{1}{2}\pa_{\m}\bar{\f}\pa_{\m}\f\left[ K_{,\f\bar{\f}}(\f+cM,\bar{\f})
+K_{,\f\bar{\f}}(\f-cM,\bar{\f})\right] \cr
& -\fracm{1}{2}\pa_{\m}\bar{\f}\pa_{\m}\f\left[ cK_{,\f\bar{\f}}(\f+cM,
\bar{\f})-cK_{,\f\bar{\f}}(\f-cM,\bar{\f})\right]\fracmm{\pa M}{\pa\f} \cr
&-\fracm{1}{2}\pa_{\m}\bar{\f}\pa_{\m}\bar{\f}\left[ 
cK_{,\f\bar{\f}}(\f+cM,\bar{\f})-cK_{,\f\bar{\f}}(\f-cM,\bar{\f})\right]
\fracmm{\pa M}{\pa\bar{\f}}~~,  \cr}\eqno(3.4a)$$
where $M(\f,\bar{\f})$ is supposed to be a solution to eq.~(2.18). Equation
(3.4a) can be rewritten to some other equivalent forms, {\it viz.}
$$L_{\rm kin.} (\f,\bar{\f})= 
-\fracm{1}{2}(\pa_{\m}\bar{\f}\pa_{\m}\f)\fracmm{\pa}{\pa\f} 
\left[ K_{,\bar{\f}}(\f+cM(\f,\bar{\f}),\bar{\f})
+K_{,\bar{\f}}(\f-cM(\f,\bar{\f}),\bar{\f})\right]$$
$$-\fracm{1}{2}(\pa_{\m}\bar{\f}\pa_{\m}\bar{\f}) \left[ 
cK_{,\f\bar{\f}}(\f+cM(\f,\bar{\f}),\bar{\f})-cK_{,\f\bar{\f}}(\f-
cM(\f,\bar{\f}),\bar{\f})\right]
\fracmm{\pa M(\f,\bar{\f})}{\pa\bar{\f}}~~,\eqno(3.4b)$$
or
$$L_{\rm kin.} (\f,\bar{\f})= 
-\fracm{1}{2}(\pa_{\m}\bar{\f}\pa_{\m}\f)\fracmm{\pa}{\pa\f}
\fracmm{\pa}{\pa\bar{\f}} 
\left[ K(\f+cM(\f,\bar{\f}),\bar{\f}) +K(\f-cM(\f,\bar{\f}),\bar{\f})\right]$$
$$+\fracm{1}{2}(\pa_{\m}\bar{\f}\pa_{\m}\f)\fracmm{\pa}{\pa\f}
\left[ cK_{,\f}(\f+cM,\bar{\f}) -cK_{,\f}(\f-cM,\bar{\f})
\right]\fracmm{\pa M(\f,\bar{\f})}{\pa\bar{\f}} $$
 $$-\fracm{1}{2}(\pa_{\m}\bar{\f}\pa_{\m}\bar{\f}) \left[ 
cK_{,\f\bar{\f}}(\f+cM(\f,\bar{\f}),\bar{\f})-cK_{,\f\bar{\f}}(\f-
cM(\f,\bar{\f}),\bar{\f})\right]
\fracmm{\pa M(\f,\bar{\f})}{\pa\bar{\f}}~~.\eqno(3.4c)$$
For instance, the first line of eq.~(3.4c) gives some K\"ahler-like terms with
the NAC-deformed (splitted) K\"ahler potential, whereas the second and third
lines of eq.~(3.4c) apparently violate both K\"ahlerian and Hermitian 
structures of the original (undeformed) NLSM when $c\neq 0$ and 
$\pa M(\f,\bar{\f})/\pa\bar{\f}\neq 0$ ({\sf cf.} the last 
ref.~\cite{chandra}).

Therefore, the NAC deformation of the K\"ahler 
kinetic terms in eq.~(1.12) amounts to a non-K\"ahlerian and 
non-Hermitian deformation of the original K\"ahlerian and Hermitian NLSM, 
which is controlled by the auxiliary field solution to eq.~(2.18). In the case
of a single chiral superfield, the deformed NLSM metric can be read off from 
eq.~(3.4). In the case of several superfields, the deformed NLSM can be read 
off from eq.~(3.1), when assuming all the $\x$-integrations to be performed 
with the auxiliary fields considered as the parameters or spectators.

 Being unable to explicitly solve eq.~(2.18) in a generic case, in the next 
sect.~4 we consider the simplest non-trivial example provided by the $CP^n$ 
NLSM with an (undeformed) K\"ahler, Hermitian and symmetric target space.

\section{NAC-deformed $CP^n$ NLSM}

The $CP^n$ (principal) NLSM is characterized by a K\"ahler potential
$$ K(\f,\bar{\f})=\a\, \ln (1 +\k^{-2}\f\bar{\f})~~,\eqno(4.1)$$
where we have used the notation 
$\f\bar{\f}\equiv \abs{\f}{}^2 \equiv \f^i\bar{\f}^i$. 
Due to a Grassmann nature of the fermionic fields, getting explicit results 
with fermions is possible. However, those results appear to be cumbersome and
not very illuminating. Therefore, we ignore fermions for simplicity in this 
section, except of the simplest $CP^1$ case.

Equations (2.18) in the $CP^n$ case read as follows:
$$ \left(\f^i-\fracmm{(1+\k^{-2}\abs{\f}{}^2)M^i}{\k^{-2}P}
\right) \left(\fracmm{1}{1+\k^{-2}\abs{\f}{}^2+c\k^{-2}P}-
\fracmm{1}{1+\k^{-2}\abs{\f}{}^2-c\k^{-2}P}\right) $$
$$ + \fracmm{2c\bar{W}_{,\bar{\f}^i}}{\a\k^{-2}} =0~~, \eqno(4.2)$$
where we have used the notation $P=M^i\bar{\f}^i$. 

Multiplying eq.~(4.2)
with $\bar{\f}^i$ gives rise to a quadratic equation on P,
$$ \fracmm{\a\k^{2}P}{(\k^2+\abs{\f}{}^2)^2-c^2P^2}+ U=0~,\eqno(4.3)$$
where we have introduced yet another notation 
$U=\bar{\f}^i\bar{W}_{,\bar{\f}^i}$. 

A solution to eq.~(4.3) is given by
$$ P=\fracmm{\a\k^{2}-\sqrt{\a^2\k^4+4U^2c^2(\k^2+\abs{\f}{}^2)^2}}{2c^2U}~~~,
\eqno(4.4)$$
where we have chosen the minus sign in front of the square root, in order to
assure the existence of the $c\to 0 $ limit. In the case of the $CP^1$ model,
eq.~(4.4) already gives a solution for $M=\bar{\f}^{-1}P$ \cite{nac2}.  

In the $CP^n$ case with $n>1$, substituting the solution (4.4) back into 
eq.~(4.2) yields the following result:
$$ M^i=\fracmm{P\bar{\f}^i}{\k^2+\abs{\f}{}^2}
- \fracmm{(\k^2+\abs{\f}{}^2)^2-c^2P^2}{\a(\k^2+\abs{\f}{}^2)}U~~.\eqno(4.5)$$

In the anti-commutative limit $c\to 0$, eq.~(4.4) yields
$$ \lim_{c\to 0} P =-\fracmm{(\k^2+\abs{\f}{}^2)^2}{\a\k^2}
\left(\bar{\f}^i\bar{W}_{,\bar{\f}^i}\right)~.\eqno(4.6)$$

In the case of vanishing anti-chiral superpotential, $\bar{W}=0$, we have 
$U=P=0$, so that the {\it bosonic} terms in the auxiliary field solution also 
vanish and, hence, there is no deformation of the bosonic $CP^n$ kinetic terms
at all. Given an arbitrary anti-chiral superpotential $\bar{W}(\bar{\F})
\neq 0 $, the deformed $CP^n$ metric is rather complicated (see e.g., the 
$CP^1$ result at the end of this section).
 
To give an explicit example of the NAC-deformed structure of the NLSM 
{\it fermionic} terms, let's consider the $CP^1$ model in the case of 
vanishing anti-chiral superpotential. Equation (2.18) now reads
$$ M \int^{+1}_{-1}d\x\,K_{,\f\bar{\f}}(\f+c\x M,\bar{\f})-\fracm{1}{2}\c^2
\int^{+1}_{-1}d\x\,K_{,\f\f\bar{\f}}(\f+c\x M,\bar{\f})=0~,\eqno(4.7)$$
whose K\"ahler function is given by eq.~(4.1). The integrations over $\x$ in 
eq.~(4.7) yield
$$ M \left[  \fracmm{-\a\k^2}{cM\bar{\f}(\k^2+\bar{\f}(\f+cM))}
+\fracmm{\a\k^2}{cM\bar{\f}(\k^2+\bar{\f}(\f-cM))} \right] \eqno(4.8)$$
$$ -\fracm{1}{2}\c^2\left[
\fracmm{\a\k^{2}}{cM(\k^2+\bar{\f}(\f+cM))^2}-
\fracmm{\a\k^{2}}{cM(\k^2+\bar{\f}(\f-cM))^2}\right]=0~,$$
which gives rise to a simple {\it cubic} equation on $M$,~\footnote{In the case
of a non-vanishing anti-chiral superpotential, one gets a quartic equation.}
$$ M^3 -(c\bar{\f})^{-2}(\k^2+\abs{\f}{}^2)^2M -\fracmm{\c^2}{c^2\bar{\f}}(\k^2
+\abs{\f}{}^2)=0~.\eqno(4.9)$$
The use of Cardano formula for the roots of a cubic equation gives us three
solutions as follows:
$$ M = \o^m\left[ -\fracmm{A_0}{2}+\sqrt{\fracmm{A^2_0}{4}+\fracmm{A^3_1}{27}}
\right]^{1/3} + 
\o^{3-m}\left[ -\fracmm{A_0}{2}-\sqrt{\fracmm{A^2_0}{4}+\fracmm{A^3_1}{27}}
\right]^{1/3}~~,\eqno(4.10)$$
where $\o=(-1+\sqrt{3})/2$, $m=0,1,2$ and 
$$ \eqalign{
A_0 & = -\fracmm{\c^2}{c^2\bar{\f}}(\k^2+\abs{\f}{}^2)~,\cr
A_1 & = -(c\bar{\f})^{-2}(\k^2+\abs{\f}{}^2)^2~.\cr}\eqno(4.11)$$

However, due to the nilpotency property of fermions, $(\c^2)^2=0$ in the 
$CP^1$ case,~\footnote{In the $CP^n$ case, we merely have 
$(\c^i\c^j)^{n+1}=0$.} eq.~(4.10) is greatly simplified to the followong
three solutions:
$$ M_0 = -\fracmm{\bar{\f}\c^2}{\k^2+\abs{\f}{}^2} \eqno(4.12a)$$
and
$$ M_{\pm}= \fracmm{\pm(\k^2+\abs{\f}{}^2)}{c\bar{\f}}+
\fracmm{\c^2\bar{\f}}{2(\k^2+\abs{\f}{}^2)}~~.\eqno(4.12b)$$ 

The solution (4.12a) is clearly the same as that in the undeformed case, 
whereas the other two solutions are singular in the undeformed limit $c\to 0$.
We are unaware of any possible physical significance of the singular solutions.

Inserting the solution (4.12a) into eqs.~(2.16) and (2.17), and using the 
nilpotency condition, $(\c^2)^2=0$, give rise to the standard (undeformed) 
four-dimensional N=1 supersymmetric $CP^1$ NLSM in components. It appears to 
coincide with the result \cite{chuo} in the $CP^1$ case, where also no
NAC-deformation of the $CP^1$ supersymmetric NLSM was discovered, though our 
NAC-deformation is different from the one considered there (see sect.~5). 

It is of interest to get an explicit NAC-deformed NLSM metric in the case of a
{\it non-vanishing} anti-chiral potential, $\bar{W}\neq 0$. Even, in the $CP^1$
case, it gives rise to a rather complicated metric, when using eqs.~(3.4), 
(4.1) and the auxiliary field solution 
$$ M =\fracmm{\a -\sqrt{\a^2+\left(2c\bar{\f}(1+\k^{-2}\f\bar{\f})
\bar{W}_{,\bar{\f}}\right)^2}}{2c^2\k^{-2}\bar{\f}^2\bar{W}_{,\bar{\f}}}
\eqno(4.13)$$
that follows from eq.~(4.4), where we have used the notation
$\bar{W}_{,\bar{\f}}=\pa\bar{W}/\pa\bar{\f}$. A straightforward calculation 
(both `by hand' and by the use of a Mathematica computer program) in the 
$CP^1$ case yields the following deformed NLSM kinetic terms:
$$ L_{\rm kin.}= -g\low{\f\f}\pa_{\m}\f\pa_{\m}\f
  -2g\low{\f\bar{\f}}\pa_{\m}\f \pa_{\m}\bar{\f}
- g\low{\bar{\f}\bar{\f}}\pa_{\m}\bar{\f}\pa_{\m}\bar{\f}~~,\eqno(4.14)$$
where
$$\eqalign{
g\low{\f\bar{\f}} =~& 
\fracmm{-\a\k^{-2}c^2\bar{\f}^2(\bar{W}_{,\bar{\f}})^2}{
\left(-\a+\sqrt{\a^2+\left(2c\bar{\f}(1+\k^{-2}\f\bar{\f})
\bar{W}_{,\bar{\f}}\right)^2}\right)
\sqrt{\a^2+\left(2c\bar{\f}(1+\k^{-2}\f\bar{\f})\bar{W}_{,\bar{\f}}\right)^2}}
~~~,\cr
g\low{\f\f} = & ~0~,\cr
g\low{\bar{\f}\bar{\f}} =~ & 
\fracmm{-2\a^{-1}c^2(1+\k^{-2}\f\bar{\f})\bar{W}_{,\bar{\f}}}{\left(
\a-\sqrt{\a^2+\left(2c\bar{\f}(1+\k^{-2}\f\bar{\f})
\bar{W}_{,\bar{\f}}\right)^2}\right)
\sqrt{\a^2+\left(2c\bar{\f}(1+\k^{-2}\f\bar{\f})
\bar{W}_{,\bar{\f}}\right)^2}} ~\times \cr
& \times \left[ 4c^2\bar{\f}^2(\bar{W}_{,\bar{\f}})^3(1+\k^{-2}\f\bar{\f}) 
\right.  \cr
& \left. +~ \a\left( \a-\sqrt{\a^2+\left(2c\bar{\f}(1+\k^{-2}\f\bar{\f})
\bar{W}_{,\bar{\f}}\right)^2}\right) (2\bar{W}_{,\bar{\f}}+\bar{\f}
\bar{W}_{,\bar{\f}\bar{\f}})\right]~.\cr}\eqno(4.15)$$
It is worth noticing that $\det g =-(g\low{\f\bar{\f}})^2$. The most apparent
feature $g\low{\f\f}=0$ is also valid in the case of a generic NAC-deformed 
NLSM (in a given parametrization). 

\section{Conclusion}

As is clear from our discussion and explicit examples, the problem of solving 
for the auxiliary fields is highly non-trivial in the context of NAC-deformed 
N=1/2 supersymmetric NLSM under consideration. Special care should be 
exercised when considering this problem with the smearing effects that should
be calculated first.

One should also distinguish between a NAC-deformation and N=1/2 supersymmetry.
Though the NAC-deformation we considered is N=1/2 supersymmetric, the former 
is {\it stronger} than the latter. Requiring merely N=1/2 supersymmetry of a
four-dimensional NLSM would give rise to much weaker restrictions on the NLSM
target space, as is pretty obvious from eqs.~(1.10) and (1.11). 

Some comments are in order about the relation between our results in this 
paper and those obtained in ref.~\cite{chuo} for the N=1/2 supersymmetric 
NAC-deformed $CP^n$ NLSM, in the absence of chiral and anti-chiral scalar 
superpotentials.

As is well known in the theory of NLSM (see e.g., ref.~\cite{nlsm}), the 
so-called quotient construction (or gauging isometries of the `flat' NLSM 
target space) can be used to represent some NLSM with homogeneous target 
spaces as the gauge theories. It was used in ref.~\cite{chuo} to construct the
NAC-deformed N=1/2 supersymmetric NLSM in four dimensions with the $CP^n$ 
target space, by combining the quotient construction and the results of 
ref.~\cite{sei} about the NAC-deformed supersymmetric gauge theories. 
Unlike the NAC-deformation of the non-gauge supersymmetric field theory (1.12)
governed by the scalar deformation parameter $c$, the NAC-deformed N=1/2 
supersymmetric field theory \cite{sei} has $C_{\a\b}$ as the deformation 
parameters, while its action is {\it not} `Lorentz'-invariant. In addition,
there are more auxiliary fields in the supersymmetric quotient construction, 
while it does not lead to any splitting or smearing of the K\"ahler potential.
As was demonstrated in ref.~\cite{chuo}, the quotient construction of the 
NAC-deformed N=1/2 supersymmetric $CP^n$ model does {\it not} modify the 
K\"ahler geometry at all, while the {\it only} new term in the deformed 
Lagrangian takes the form \cite{chuo}
$$ g_{p\bar{q}}g_{r\bar{s}}C^{\a\b}(\s^{\m\n})\du{\b}{\g}\c^p_{\a}\c^r_{\g}
\pa_{\m}\bar{\f}^{\bar{q}}\pa_{\n}\bar{\f}^{\bar{s}}~~,\eqno(5.1)$$
where $ g_{p\bar{q}}(\f,\bar{\f})$ is the Fubini-Study metric associated with
the K\"ahler potential (4.1).
As is clear from eq.~(5.1), this deformation is linear in $C^{\a\b}$, while
it is not `Lorentz'-invariant. The quotient construction itself is also limited
to the homogeneous NLSM, and it does not allow a scalar superpotential.

Our approach to the NAC-deformed NLSM is very general, while the NAC 
deformation of a K\"ahler potential is controlled by the auxiliary fields 
$M^i$ entering the deformed K\"ahler potential in the highly non-linear way. 
In the absence of an anti-chiral superpotential, when requiring the smooth 
undeformed limit, the bosonic part of the auxiliary field solution vanishes, so
that we are left with the undeformed bosonic K\"ahler potential as well. 
However, when the auxiliary fields are eliminated, the structure of the 
fermionic terms in our approach is apparently not the same as that of 
ref.~\cite{chuo}, even modulo possible field redefinitions.

At the same time, the N=1/2 supersymmetry transformation laws are not modified
by a NAC-deformation, both in our approach and in ref.~\cite{chuo}. Our 
conclusion is that the two methods give rise to the inequivalent N=1/2 
supersymmetric extensions of the K\"ahler NLSM. It does not seem to be very
surprising because there are many N=1/2 supersymmetric extensions of the NLSM
kinetic terms.


\begin{thebibliography}{99}

\bibitem{sei} N. Seiberg, JHEP {\bf 0306} (2003) 010 [hep-th/0305248]
\bibitem{chandra} B. Chandrasekhar and A. Kumar, JHEP {\bf 0403} (2004) 013 
[hep-th/0310137];\\
B. Chandrasekhar, Phys.Rev. {\bf D70} (2004) 125003 [hep-th/0408184];\\
B. Chandrasekhar, {\it N=2 $\s$-model action on  non(anti)commutative 
superspace}, hep-th/0503116
\bibitem{chuo} T. Inami and H. Nakajima, Progr. Theor. Phys. {\bf 111} (2004)
961 [hep-th/0402137]
\bibitem{buch} O.D. Azorkina, A.T. Banin, I.L. Buchbinder and N.G. Pletnev,
{\it Generic chiral superfield model on nonanticommutative $N=\ha$ superspace},
 hep-th/0502008
\bibitem{nac1} T. Hatanaka, S. V. Ketov, Y. Kobayashi and S. Sasaki, 
Nucl. Phys. {\bf B716} (2005) 88 [hep-th/0502026]
\bibitem{spain} L. Alvarez-Gaume and M. Vazquez-Mozo, 
{\it On nonanticommutative N=2 sigma-models in two dimensions}, hep-th/0503016
\bibitem{nbi} T. A. Ryttov and F. Sannino, {\it Chiral models in 
nonanticommutative N=1/2 four-dimensional superspace}, hep-th/0504104 
\bibitem{nac2} T. Hatanaka, S. V. Ketov and S. Sasaki, {\it Summing up
non-anti-commutative K\"ahler potential}, hep-th/0504191
\bibitem{zumino} B. Zumino, Phys. Lett. {\bf B87} (1979) 203
\bibitem{nlsm} S. V. Ketov, {\it Quantum Non-linear Sigma-Models}, 
Springer-Verlag, Heidelberg, 2000
\bibitem{wb} J. Wess and J. Bagger, {\it Supersymmetry and Supergravity}, 
Princeton University Press, 1992
\bibitem{kgn} S. J. Gates, Jr., S. V. Ketov and H. Nishino, 
Nucl. Phys. {\bf B393} (1993) 149 [hep-th/9207042]
\bibitem{shif} A. Gorsky and M. Shifman, Phys. Rev. {\bf D71} (2005) 025009
[hep-th/0410099] 
\bibitem{arg} L. G. Aldrovandi, F. A. Schaposnik and G. A. Silva, {\it 
Non(anti)commutative superspace with coordinate-dependent deformation},
hep-th/0505217 
\bibitem{konishi} K. Konishi, Phys. Lett. {\bf B135} (1984) 439 
\bibitem{five} R. Argurio, V. L. Campos, G. Ferretti and R. Heise, Phys. Rev.
{\bf D67} (2003) 065005 [hep-th/0210291];\\
A. Brandhuber, H. Ita, H. Nieder, Y. Oz and  C. R\"omelsberger,
Adv. Theor. Math. Phys. {\bf 7} (2003) 269 [hep-th/0303001] 
\bibitem{vy} G. Veneziano and S. Yamkielowicz, Phys. Lett. {\bf B113} (1982) 
231
\bibitem{chu} C.-S. Chu and T. Inami, {\it Konishi anomaly and central 
extension in N=1/2 supersymmetry}, hep-th/0505141. 
\end{thebibliography}
\end{document}